\documentclass[12pt]{iopart}
\usepackage{iopams,graphicx,breqn}
\begin{document}
\title{Sudden violation of the CHSH inequality in a two qubits system}
\author{Laura Mazzola}
\address{Department of Physics and Astronomy, University of Turku,
FI-20014 Turun yliopisto, Finland} \ead{laumaz@utu.fi}

\newcommand{\ket}[1]{\displaystyle{|#1\rangle}}
\newcommand{\bra}[1]{\displaystyle{\langle#1|}}
\newcommand{\al}{\alpha}
\newcommand{\Om}{\Omega}
\newcommand{\om}{\omega}
\newcommand{\D}{\Delta}
\newcommand{\G}{\Gamma}
\newcommand{\g}{\gamma}
\newcommand{\si}{\sigma}
\newcommand{\la}{\lambda}

\begin{abstract}
I study the dynamics of the violation of the CHSH inequality for two
qubits interacting with a common zero-temperature non-Markovian
environment. I demonstrate sudden violation of the inequality for
two qubits initially prepared in a factorized state. Due to the
strong coupling between the qubits and the reservoir, the dynamics
is characterized by numerous sharp revivals. Furthermore I focus on
a more realistic physical system in which the spontaneous emission
for the qubits is taken into account. When including spontaneous
emission even for small decay parameters, revivals in the violation
are heavily damped out. If the decay rates exceed a certain
threshold, the inequality turns out to be always satisfied.
\end{abstract}

\maketitle

\section{Introduction}

The interaction of a quantum system with an environment is generally
seen as source of decoherence, leading to the loss of quantum
properties and the appearance of the classical world \cite{Breuer}.
In fact, due to the creation of correlation with the environment,
the evolution of the quantum system is not unitary any more. If the
system is prepared in an entangled state, such entanglement is very
likely to be lost \cite{YuScience}. However, environments can also
create correlations among quantum systems. For example, the
interaction of a bipartite system with a common reservoir creates
correlations between the parts \cite{Benatti}, irrespective of the
existence of a direct coupling between them \cite{Ficek,FicekESB}.
In a shared environment highly entangled long-living states, or even
sub-radiant decoherence free states can exist \cite{Tanas}. Many
studies have demonstrated the existence of non-zero stationary
entanglement in systems of two qubits or two harmonic oscillators
sharing the same reservoir \cite{Paz}. The ability to create
correlations persists if the common reservoir has non zero
temperature \cite{Braun}, and is increased by memory effects in
non-Markovian reservoirs \cite{Maniscalco,Francica,Mazzola}.

In the last decade many efforts have been done to understand the
dynamics of entanglement and quantum correlations in systems which
are good candidates for applications in quantum information theory
and technology. However, not all the protocols in quantum
information theory rely on entanglement. On the contrary, in many
cases entanglement is simply not enough and non-local properties,
expressed by the violation of a Bell inequality, are required.

Here I study the non-local properties, namely the violation of the
CHSH inequality, of a system of two qubits interacting with a common
non-Markovian reservoir. I consider a factorized state of the qubits
with one excitation, and study, as a function of time and of the
qubit-cavity coupling, the sudden violation of CHSH inequality. The
conditions for which the violation is maximum are identified.
Afterwards I add another dynamical ingredient and study how the
dynamics of the CHSH violation gets modified when the spontaneous
emission of the qubits is included.

\section{The model}

The system under investigation comprises two qubits interacting with
a common zero-temperature bosonic reservoir. The Hamiltonian of the
system in rotating-wave approximation, and in units of $\hbar$, is
$H=H_{0}+H_{\mathrm{int}}$,
\begin{equation}
   H_{0}=\om_{0}(\si_{+}^{(1)}\si_{-}^{(1)}+\si_{+}^{(2)}\si_{-}^{(2)})
   +\sum_{k}\om_{k}a_{k}^{\dag}a_{k},\label{H0bare}
\end{equation}
\begin{equation}\label{Hintbare}
   H_{\mathrm{int}}=(\al_{1}\si_{+}^{(1)}+\al_{2}\si_{+}^{(2)})\sum_{k}g_{k}a_{k}+\mathrm{h.c. },
\end{equation}
where $\si_{\pm}^{(1)}$ and $\si_{\pm}^{(2)}$ are the Pauli raising
and lowering operators for qubit 1 and 2 respectively, $\om_{0}$ is
the Bohr frequency of the two identical qubits, $\al_{1}$ and
$\al_{2}$ are dimensionless environment-qubit coupling constants,
$a_{k}$ and $a_{k}^{\dagger}$, $\om_{k}$ and $g_{k}$ are the
annihilation and creation operators, the frequency and the coupling
constant of the field mode $k$, respectively.

In the following I assume that the two qubits interact resonantly
with a lossy cavity, so I choose a Lorentzian spectral distribution
to describe the properties of the environment,
\begin{equation}
   J(\omega)=\frac{W^2}{\pi}\frac{\lambda}{(\om-\om_{0})^2+\lambda^2},
\end{equation}
where $\lambda$ is the width of the spectral distribution describing
the cavity losses, and $W$ in the limit of ideal cavity (when
$\lambda\rightarrow0$) is proportional to the vacuum Rabi frequency
$R$ through $W=R/\al_{T}$ with $\al_{T}=(\al_{1}^2+\al_{2}^2)^{1/2}$
the collective coupling constant.

The dynamics of two qubits interacting with a common
Lorentzian-structured reservoir has been studied in Ref.
\cite{Maniscalco} for the case of one excitation, and in Ref.
\cite{Mazzola} for a generic state of two identically coupled
qubits. Since here I consider the dynamics of a factorized state
with one excitation of the form
$\ket{\Psi(t)}=c_{1}(t)\ket{10}\ket{0}_{E}+c_{2}(t)\ket{01}\ket{0}_{E}+\sum_{k}c_{k}(t)\ket{00}\ket{1_{k}}_{E}$
I am going to use the model of Ref. \cite{Maniscalco}. There the
dynamics of the qubits is expressed in the basis of super-radiant
and sub-radiant states
\begin{equation}\label{Psim}
\ket{\psi_{-}}=r_{2}\ket{10}-r_{1}\ket{01},
\end{equation}
\begin{equation}\label{Psip}
\ket{\psi_{+}}=r_{1}\ket{10}+r_{2}\ket{01},
\end{equation}
where the relative coupling strengths $r_{1}=\al_{1}/\al_{T}$ and
$r_{2}=\al_{2}/\al_{T}$ have been introduced ($r_{1}^2+r_{2}^2=1$).

Considering the dynamics in such a basis is particularly convenient
since the sub-radiante state does not evolve in time. Thus the
evolution of the amplitudes of the first and second qubits is just
\begin{equation}
c_{1}(t)=r_{2}\beta_{-}+r_{1}\beta_{+}E(t),
\end{equation}
\begin{equation}
c_{2}(t)=-r_{1}\beta_{-}+r_{2}\beta_{+}E(t),
\end{equation}
with $\beta_{\pm}=\bra{\psi_{\pm}}\psi_{0}\rangle$ and
\begin{equation}
E(t)=e^{-\lambda t/2}[\cosh(\Omega
t/2)+\frac{\lambda}{\Omega}\sinh(\Omega t/2)],
\end{equation}
where $\Omega=\sqrt{\lambda^2-4R^2}$.

In order to evaluate the time evolution of the violation of the CHSH
Bell inequality I use the expression of Ref. \cite{BellomoNLE},
which allows one to express the maximum of the Bell function (by an
appropriate choice of angles) as function of the two-qubit density
matrix elements. In the $\{\ket{11},\ket{10},\ket{01},\ket{00}\}$
basis of the qubits, the maximum of the Bell function reads as
follows
\begin{equation}
B=2 \max_{i,j}\{u_{i}+u_{j}\}^{1/2}
\end{equation}
with
\begin{eqnarray}
u_{1}=(\rho_{11}+\rho_{44}-\rho_{22}-\rho_{33})^2,\nonumber\\
u_{2}=4(|\rho_{23}|^2+|\rho_{14}|^2),\\
u_{3}=4(|\rho_{23}|^2-|\rho_{14}|^2).\nonumber
\end{eqnarray}
where $\rho_{22}=|c_{1}(t)|^2$, $\rho_{33}=|c_{2}(t)|^2$,
$\rho_{23}=c_{1}(t)c_{2}^{*}(t)$ and
$\rho_{44}=1-|c_{1}(t)|^2-|c_{2}(t)|^2$. For the particular initial
state considered I have $\rho_{11}=0$ and $\rho_{14}=0$.

\section{CHSH violation dynamics}
Here I investigate the CHSH inequality violation for two qubits
prepared initially in the state
\begin{equation}\label{Psi0}
\ket{\psi(0)}=\ket{10}.
\end{equation}
I consider a cavity characterized by a high but experimentally
feasible quality factor (the width is $\lambda=10^{-1}\ \gamma_0$,
with $\gamma_{0}$ the Markovian decay rate of the qubits), and
strong coupling conditions. I set the parameter $S=R/\lambda=10$,
indicating the strength of the coupling \cite{Maniscalco}.

\begin{figure}[!]
\begin{center}
\includegraphics[width=12cm]{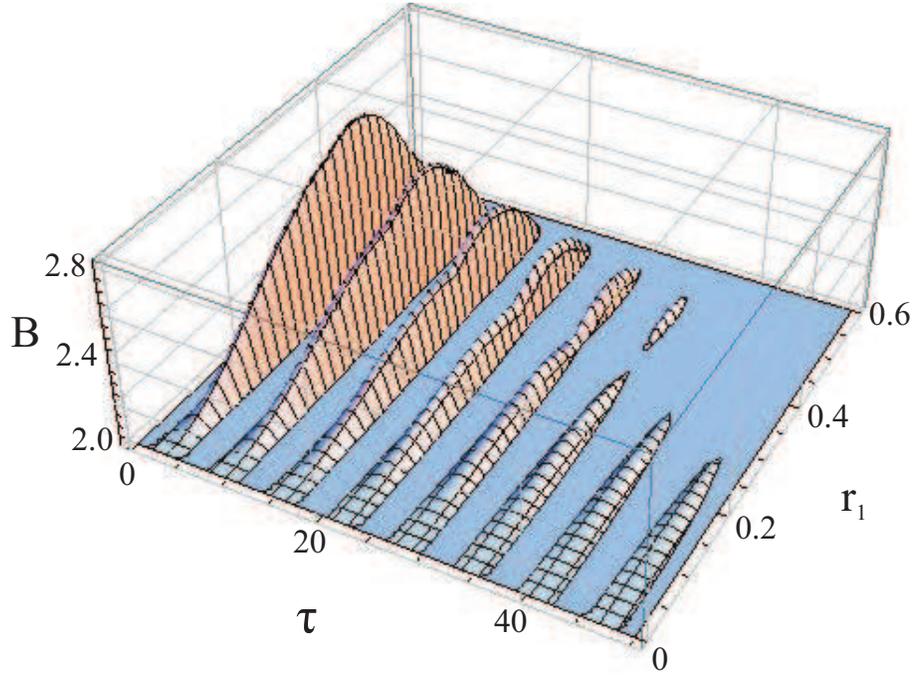}
\caption{\label{fig:VNCBell2}(Color online) Dynamics of the CHSH
inequality violation in a common Lorentzian structured reservoir as
a function of scaled time and relative coupling strength for two
atoms prepared in a factorized state $\Psi_{0}=\ket{10}$. No
spontaneous emission by the qubits is included.}\label{fig:CHSH}
\end{center}
\end{figure}

In figure \ref{fig:CHSH} I plot the CHSH violation as a function of
scaled time ($\tau=\lambda t$) and of the relative coupling strength
$r_{1}$. During the evolution the system passes through highly
entangled states, strongly violating the CHSH inequality. Cycles of
birth and death of non-locality follow one after the other, till for
long times the inequality is finally satisfied. The sudden
appearance of the violation is consequence of the shared reservoir,
providing an indirect coupling between the qubits. Such a
reservoir-mediated interaction plays certainly a role also in the
revivals of non-locality, which are also caused by the memory
effects of the non-Markovian reservoir.

I notice that sudden violation of the CHSH inequality appears up to
a certain value of the relative strength of the coupling. In
particular for the chosen coupling conditions no violation occurs
when $r_{1}>0.6$. For stronger coupling conditions the region of
violation in the $r_{1}$ space of parameters widens, nevertheless it
never reaches $1/\sqrt{2}$. In fact for $r_{1}=r_{2}=1/\sqrt{2}$ the
qubits are symmetrically coupled to the cavity, and entanglement and
non-local properties in general cannot be created out of the
factorized state in Eq. \eref{Psi0}.

\section{Including spontaneous emission}

In this section I study how the dynamical violation of the CHSH
inequality is modified when including spontaneous emission for the
two qubits. In Ref. \cite{MazCross} I have studied the entanglement
time evolution of two entangled qubits interacting with the same
Lorentzian structured reservoir (leaky cavity) and emitting
independently outside the cavity due to spontaneous emission. By
means of Eqs. (1) and (2) of Ref. \cite{MazCross} I have solved the
dynamics in the case of qubits having the same transition frequency,
equally and resonantly coupled with the cavity. Here I generalize
that master equation to include the case of different couplings
between the qubits and the cavity. So the master equation describing
the dynamics of our system (with some renaming of parameters)
becomes:
\begin{eqnarray}\label{ME}
\frac{\partial\tilde{\rho}}{\partial t}=-\imath[H,\tilde{\rho}]
-\lambda(a^{\dag}a\tilde{\rho}+\tilde{\rho}a^{\dag}a-2a\tilde{\rho}a^{\dag})\nonumber\\
-\frac{\g_{1}}{2}(\si_{+}^{(1)}\si_{-}^{(1)}\tilde{\rho}+\tilde{\rho}\si_{+}^{(1)}\si_{-}^{(1)}-2\si_{-}^{(1)}\tilde{\rho}\si_{+}^{(1)})\nonumber\\
-\frac{\g_{2}}{2}(\si_{+}^{(2)}\si_{-}^{(2)}\tilde{\rho}+\tilde{\rho}\si_{+}^{(2)}\si_{-}^{(2)}-2\si_{-}^{(2)}\tilde{\rho}\si_{+}^{(2)}),
\end{eqnarray}
with
\begin{equation}
H=\al_{T}W[(r_{1}\si_{+}^{(1)}+r_{2}\si_{+}^{(2)})a+\mathrm{h.c. }].
\end{equation}
where $\si_{\pm}^{(1)}$ and $\si_{\pm}^{(2)}$, $\al_{t}$, $r_{1}$
and $r_{2}$, $\lambda$ and $W$ have been defined above, while $a$
and $a^{\dagger}$ are the annihilation and creation operators for
the cavity mode and $\g_{1/2}$ are the spontaneous emission rates
for the two qubits. For the sake of simplicity I assume
$\g_{1}=\g_{2}=\g_{S}$.

\begin{figure}[!]
\begin{center}
\includegraphics[width=12cm]{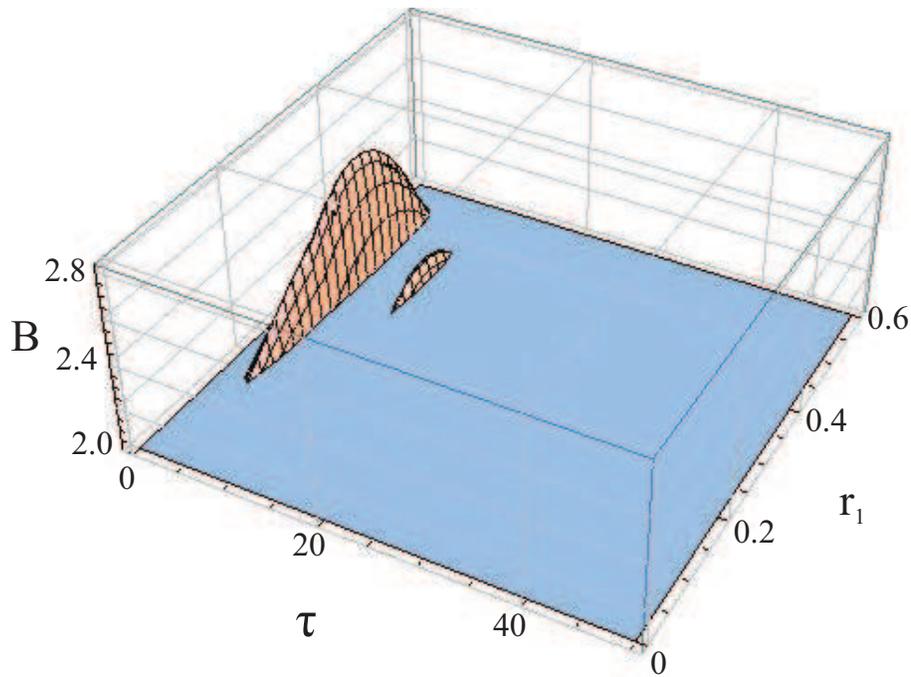}
\caption{\label{fig:VNCBell2}(Color online) Dynamics of the CHSH
inequality violation in a common Lorentzian structured reservoir as
a function of scaled time and relative coupling strength for two
atoms prepared in a factorized state $\Psi_{0}=\ket{10}$. Here
spontaneous emission is included and the spontaneous emission decay
rate of the qubits are equal to $(1/50) \gamma_{0}$, i.e., 50 times
smaller than the vacuum Rabi frequency
$R=\al_{T}W$.}\label{fig:CHSHspont}
\end{center}
\end{figure}

As in the previous section I consider the sudden appearance of CHSH
violation for two qubits prepared in the state
$\ket{\psi(0)}=\ket{10}$. I study the dynamics for the same strong
coupling conditions $S=R/\lambda=10$, in particular for the same
width of the spectral distribution, and vacuum Rabi frequency. The
two-qubit spontaneous emission decay rates are set equal to
$\gamma_{S}/\gamma_{0}=1/50$, therefore they are 50 times smaller
than the vacuum Rabi frequency $R$. In figure \ref{fig:CHSHspont} I
show the CHSH violation as a function of time and of the relative
coupling parameter $r_{1}$. The effect of spontaneous emission is
apparent: the long series of sharp revivals is dramatically damped
out. For the chosen spontaneous emission parameter only one small
revival is present. The amount of violation of the CHSH inequality
reduces with increasing decay rate, and beyond a certain threshold
value the CHSH inequality is always satisfied. For the coupling
conditions chosen such a threshold parameter is $\gamma_{S}=1/9 \
\gamma_{0}$.

\section{Conclusion}

I have investigated the time evolution of the violation of the CHSH
inequality for a system of two qubits interacting with a common
Lorentzian structured reservoir. The reservoir-induced correlations
between the qubits drive the evolution of an initial factorized
state as $\ket{\psi(0)}=\ket{10}$ towards highly entangled states,
strongly violating the CHSH inequality. After cycles of birth and
death of non-locality, progressively damped out, the CHSH inequality
becomes always satisfied. As a second step I consider a more
realistic system in which spontaneous emission of the two qubits is
taken into account. The time evolution of the CHSH violation well
describes the dramatic changes that the introduction of spontaneous
emission in this model brings to the dynamics of the system. Birth
of non-local properties appears only if the spontaneous emission
decay rate is below a certain threshold, even in that case
oscillations are strongly damped and the inequality becomes
permanently satisfied much earlier.

The dynamics of the CHSH violation was studied for a system of two
non interacting initially entangled qubits in two independent
non-Markovian reservoirs \cite{BellomoNLE,Bellomo}. There, death and
revivals of the violation have been observed due to the memory
effects of the reservoirs. In that case however, correlations
between the qubits cannot be created starting from a factorized
state, as a consequence an initially separable state cannot evolve
into an entangled one, and possibly violating the CHSH inequality.

\ack

I would like to thank S. Maniscalco and J. Piilo for enlightening
discussions, G. Compagno and his group in Palermo for the kind
hospitality and useful suggestions. Finally I thank M. Ehrnrooth
Foundation for financial support.

\end{document}